\documentclass[a4paper,11pt]{article}
\usepackage{latexsym}
\usepackage{graphicx}

\author{E. Breschi, G. Kazakov, B. Matisov, G. Mileti \\
Laboratoire Temps-Fr\'equence - IMT, University of Neuch$\hat{a}$tel,\\
  rue A.-L.-Breguet 1, CH-2000 Neuch$\hat{a}$tel, Switzerland \\
  St. Petersburg State Polytechnic University, Polytechnicheskaya 29\\
  St. Petersburg, 195251 St. Petersburg, Russia
evelina.breschi@unine.ch}

\title{Study of lin$\|$lin CPT for application in vapour cell atomic clocks}

\begin{document}
\maketitle

\section{Abstract}
We evaluate the use of Coherent Population Trapping (CPT) excited with parallely polarized laser fields in vapour cell atomic clocks. We study the resonance shape, the discriminator slope and signal-to-noise ratio as a function of relevant parameters such as the laser frequency detuning and the applied magnetic field. We show that a stability of $6 \cdot 10^{-12}$ $\tau^{-1/2}$ may be reached in a compact system using a modulated VCSEL.

\section{Introduction}
Contrarily to the microwave-optical double resonance scheme, Coherent Population Trapping (CPT) 
does not require a microwave cavity, which may be an advantage in compact or miniature atomic
clocks \cite{Knappe, Vanier05}. CPT usually suffers from smaller signal contrast, 
even though dif\-fe\-rent techniques have been recently proposed to overcome this potential limitation [3 - 6].

In this communication we present our study on the use of
CPT in vapour cell frequency standards. We have investigated the case of $^{87}$Rb D$_1$ when a longitudinal magnetic field is applied and
the CPT is excited by a linearly polarized, unidirectional and multi-frequency light field $($the so called lin$\|$lin CPT$)$ [7 - 10]. Here below the experimental setup is shortly described. In section two we present the signal shape for different values of magnetic field and we discuss the maximum contrast achievable in our configuration. 
In section three, the role of the laser frequency detuning, the noise measurements and determination of the optimal lock-in detection frequency are discussed. Finally, we predict the achievable clock frequency stability, basing our considerations on the presented signal-to-noise measurements.

\begin{figure}[htbp]
\centering\includegraphics[width=7cm]{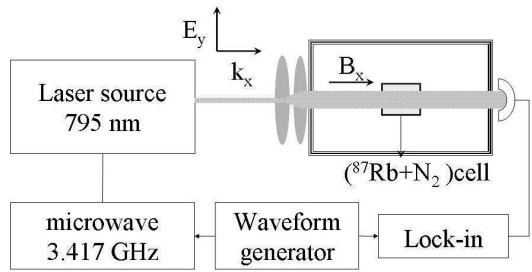}
\caption{The experimental setup used for lin$\|$lin CPT.}
\label{fig:Setup}
\end{figure}

Figure \ref{fig:Setup} shows our setup.
The Rb atoms are confined in a cylindrical glass cell (volume 1 cm$^3$) placed in a
cylindrical magnetic shield (volume 40 cm$^3$). A longitudinal
magnetic field (B$_l$) is provided by a solenoid placed inside the magnetic shield, B$_l$ is parallel to the laser beam propagation
vector.  The experiments were performed by using two cells containing either $0.5$ or $1.5$
kPa of nitrogen as buffer gas; higher buffer gas pressure in the cell are not suitable because of the suppression of quadrupole transitions due to the excited state mixing induced by the buffer gas. Remark that this intrinsic characteristic prohibits the application of the lin$\|$lin CPT in miniature-clocks unless wall-coating is used, because in the mini-cells (volume $<1$ mm$^{3}$) a buffer gas pressure of several tens of kPa is necessary.

The multi-frequency optical spectrum is generated via modulation of the
injection current of a single mode VCSEL chip emitting around 795
nm \cite{Affolderbach-VCSEL}. The VCSEL used in this
experiment has a linewidth of (100$\pm$20) MHz, measured from a 
heterodyne beatnote.
During the ex\-periments the VCSEL was modulated at the frequency of 3.417 GHz
(i.e. half of ground state hyperfine separation in $^{87}$Rb) with a power
of 10 dBm. We estimated the spectral properties of the modulated VCSEL by
recording the $^{87}$Rb one-photon absorption for different modulation
frequencies and amplitudes. We did not observe
effects related to amplitude modulation.
The phase modulation index was about
1.8, corresponding to about 70$\%$ of the total laser power in the 1st
order side-bands, which are used for CPT preparation. The remaining
power is di\-stri\-bu\-ted in the carrier, 2nd and 3rd order side-bands, and essentially does not contribute to the CPT signal but
increases the dc "background" level on the photo-detector, induces AC Stark shift and contributes to one-photon absorption processes .  

\section{Degenerate and non-degenerate CPT}

\begin{figure}[htbp]
\centering\includegraphics[width=7cm]{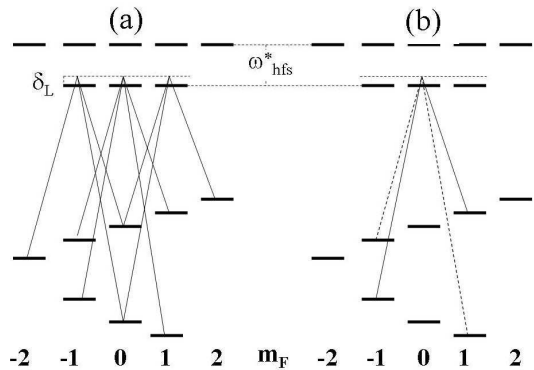}
\caption{Interaction scheme for lin$\|$lin hyperfine CPT. (a)
  represents all $\sigma^+$ and $\sigma^-$ allowed transitions; while (b) represents
  the transitions contributing to the resonance approaching the B$_l=0$ resonance frequency.} 
\label{fig:Int}
\end{figure}

Figure \ref{fig:Int} shows the relevant atomic levels and possible transitions for the $^{87}$Rb D$_1$ line induced by the 
$\sigma^+$ and $\sigma^-$ components of the linearly polarized light fields. 
We focus our attention on the group of transitions from
the ground state F$_g=1,2$ towards the excited state F$_e$ = 1 because the maximum contrast is expected in this
configuration \cite{Taichenacev-1}. For the application in frequency standards we take into account the resonances less sensitive to the magnetic field, i.e. the CPT signal occurring between ground state Zeeman sublevels with m$=0$ and m$=\pm1$. Remark that, in the case of linear polarization, the $\sigma^{+}-\sigma^{+}$ and $\sigma^{-}-\sigma^{-}$ groups of transitions starting from the ground state Zeeman sublevels with m$=0$, give rise to two orthogonal CPT states which interfere destructively 
and do not contribute to the CPT signal. As a consequence, we have to consider only the two CPT states created by the coherent superposition of $|$F$_g=1$, m$_F= \pm1>$ and
$|$F$_g=2$, m$_F=\mp 1>$, represented by the dashed and solid line in figure \ref{fig:Int}(b), respectively. These two CPT states are symmetrically split by the magnetic field with a
factor of $\pm$28 Hz $\mu$T$^{-1}$ determined by the nuclear g-factor, and are both
shifted by a factor of 0.043 Hz $\mu$T$^{-2}$.
In principle, the sensitivity to the magnetic field makes these CPT states not suitable for atomic clocks. However, their supposed high contrast drove several studies of their metrological properties [7 - 10]. The first observation of the CPT resonance splitting due to the nuclear g-factor, is reported for Cs atoms in reference \cite{Knappe-99}. In \cite{Kazakov05}, we have presented a detailed theoretical analysis of the lin$\|$lin CPT lineshape and proposed to use as frequency discriminator the maximum of absorption between the two CPT peaks when degeneracy is removed (so-called pseudo-resonance). Zibrov and co-workers \cite{Zibrov} accounted for measuring the main metrological characteristic of the degenerate lin$\|$lin CPT and pseudo-resonance. In their experimental setup, the multi-frequency spectrum was obtained by modulating the current of a diode laser which was injection locked by the radiation of a single mode ECDL. Therefore, our study constitutes the first experimental evaluation of the achievable clock frequency stability with lin$\|$lin CPT in a single laser compact set-up.

\begin{figure}[htbp]
\centering\includegraphics[width=7cm]{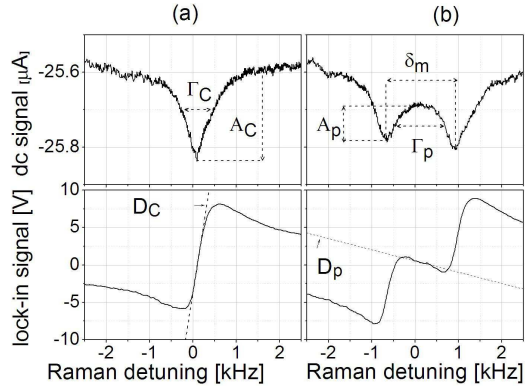}
\caption{Degenerate CPT (B$_{l}< 8$ $\mu$T) and pseudo-resonance (B$_{l} \approx 30$ $\mu$T) signals and their correspondent
  discriminator signals obtained by phase-sensitive detection.} 
\label{fig:dis}
\end{figure}

Figure \ref{fig:dis} shows typical degenerate CPT and pseudo-resonance signals obtained in our ex\-periments: in figure \ref{fig:dis}(a), the relative magnetic shift ($\delta_m$) between the
two CPT states is smaller than their linewidth ($\Gamma_C$). In this case the two CPT states are degenerate and their signals merge. In figure \ref{fig:dis}(b), the B$_l$ applied is such as $\delta_m>\Gamma_C$, as a consequence the CPT degeneracy is removed and the pseudo-resonance signal appears.
Note that $\Gamma_{C}$ determines the B$_l$ to apply for obtaining the maximum
pseudo-resonance amplitude, which corresponds to the amplitude of
the non-degenerate CPT (A$_p$= A$_C$). Considering that, in our experimental conditions, the CPT signal is well
approximated by a Lorentzian function, A$_p$= A$_C$ means $\Gamma_{p}=1.3$ $\Gamma_{C}$. Since the
amplitude of the non-degenerate CPT is smaller than the amplitude of the degenerate CPT, the optimal discriminator slope
for pseudo-resonance will be approximately 3 times worse than the degenerate CPT discriminator slope.

In Figure \ref{fig:dis} the CPT signal contrast, defined as the ratio between the CPT amplitude and DC level on the photodetector, is less than $1\%$. By optimizing the experimental parameters (buffer gas pressure, cell temperature, laser diameter) in our setup, a maximum contrast of $3 \%$ has been recorded. These values are noticeably smaller than the contrast obtained with the optical microwave double resonance. The main limiting factor to reach a higher CPT contrast is connected with the choice of creating the multi-frequency spectrum by a direct modulation of the VCSEL injection current. Whith this technique, when the amplitude of the $1^{st}$-order sidebands is maximized, the $30\%$ of the total laser power is distributed in the carrier and higher order sidebands. Considering that the distance between two consecutive frequencies in the optical spectrum is about $3.4$ GHz and the typical Doppler linewidth is about $\Gamma_D \approx 0.5$ GHz; in CPT resonance conditions, most of the power of the carrier and higher order sidebands pass trough the cell without being absorbed by $^{87}$Rb atoms, increasing significantly the background photodetector signal. Let us consider, for instance, the case of figure \ref{fig:dis}: the total laser power was about $200$ $\mu$W of which $60$ $\mu$W contributed mainly to the photodetector background signal. Whith the photodetector sensitivity of about $0.4$ A W$^{-1}$, the background signal due to the carrier and the higher order sidebands is roughly $20$ $\mu$A. In principle, a degenerate lin$\|$lin CPT contrast up to $10\%$ may be reached using a pure bichromatic field.

\section{Detection signal and noise: expected short term stability}
\begin{figure}[htbp]
\centering\includegraphics[width=7cm]{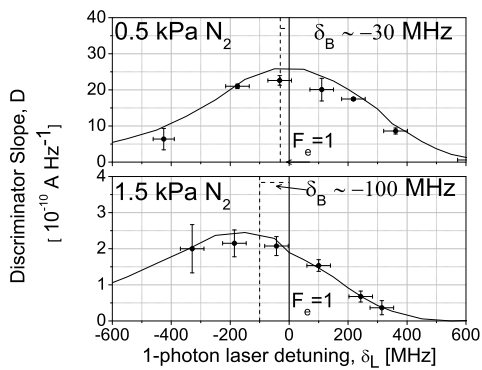}
\caption{Experimental (dot) and theoretical (line) D versus
  laser detuning ($\delta_L$) for two cells filled with different buffer gas pressure,
  $0.5$ and $1.5$ kPa, respectively. $\delta_L$ =0 refers to the absorption F$_g=1,2
  \rightarrow$ F$_e=1$ in an evacuated $^{87}$Rb cell.The shift of the D maximum depends on both the
pressure shift (dotted line) and broadening of the D$_1$ line.}
\label{fig:DisVDet}
\end{figure}

A comparison of lin$\|$lin CPT for the laser locked to either F$_e$=1 or F$_e$=2 group of transitions has been presented in \cite{Taichenacev-1}, where the authors showed that the group of transitions
towards F$_e$=1 results in larger A$_C$ and narrower $\Gamma_{C}$.
In order to determine the optimal laser detuning for frequency standard applications, we studied D$\sim$(A$_{C} \cdot \Gamma_{C}^{-1}$) versus the laser detuning ($\delta_L$) around the group of transitions towards F$_e$=1 (figure \ref{fig:DisVDet}). 
We observe that the laser detuning corresponding to the maximum of D depends on the buffer gas pressure in the cell. By comparing experimental and theoretical results, we pointed out that both the pressure shift and broadening of the $^{87}$Rb D$_1$ influence the lin$\|$lin CPT signal. In the case of a cell containing a N$_2$ pressure of $1.5$ kPa, for instance, the maximum value of D corresponds to a detuning of about $-200$ MHz, while the pressure shift is only $-100$ MHz (lower plot in figure \ref{fig:DisVDet}). We interpret this difference as a results of the destructive quantum interference between CPT states occurring when both levels F$_e=1$ and F$_e=2$ are excited simultaneously. If the laser is tuned resonant to F$_e$=1, the influence of the transition towards F$_e$=2 is due to the pressure broadening of the hyperfine transitions and/or the laser linewidth. Further experiments are in progress to quantify the dependence of the quantum interference on the experimental parameters.

The short term stability of an atomic clock depends on the parameters of the 
reference signal (A$_C$, $\Gamma_C$), the resonance frequency ($\omega_{hfs}$) and the detection noise ($N$). 
It can be evaluated by the Allan deviation, 
$\sigma_{A} =0.2 \cdot (\Gamma_C \cdot A_C^{-1}) \cdot N \cdot \omega_{hfs}^{-1} \cdot \tau ^{-1/2}$ \cite{Vanier-Allan}.

\begin{figure}[htbp]
\centering\includegraphics[width=7cm]{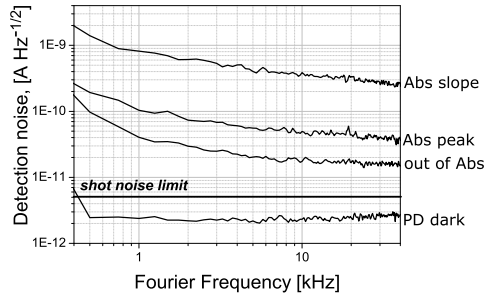}
\caption{Detection noise (N) measured for the dark photo-detector, for the
  laser field tuned out, tuned to the maximum and to the slope of the $^{87}$Rb D$_1$ line.}  
\label{fig:noise}
\end{figure}

The noise power spectrum has been measured for different the laser detuning ($\delta_L$) in our setup, using
 the cell filled with $^{87}$Rb and 1.5 mbar of N$_2$ (figure \ref{fig:noise}). We observed that the
detection noise level depends on the laser frequency detuning: when the laser is resonant with the Rb absorption the 
detection noise level increases significantly, indicating that the dominant contribution to the detection noise is the laser FM-to-AM conversion \cite{Camparo-98}. 
At higher frequency the detection noise is noticeably reduced: a factor 20 when changing $\omega$ from $0.4$ kHz to $40$ kHz for the laser tuned to the absorption peak. As a consequence it could be convenient for CPT-based atomic clocks to work with a lock-in modulation frequency ($\omega$) up to several tens of kHz. In this case the $\omega$ is larger than the linewidth of the monitored signal \cite{Merimaa-03, 06-Gerginov}, which can not be implemented in the micro-wave optical double resonance clock scheme as presently used. On the contrary, in reference \cite{Merimaa-03}, the authors showed that it is possible realize a CPT clock using $\omega \approx 10 \cdot \Gamma_C$.  In our experimental conditions we verified that $\omega$ up to 40 kHz can be used for phase-sensitive detection when $\Gamma_C$ is about $1$ kHz. We record a reduction of the discriminator signal amplitude by a factor $5$ when we changed $\omega$ from $0.4$ kHz to $40$ kHz. Therefore the signal-to-noise ratio results $4$ times larger at $\omega=40$ kHz than $\omega=0.4$ kHz.

In table \ref{tab1} we show the expected optimal short term stability for the cell containing 1.5 mbar of N$_2$. The notable
difference between the theoretical and experimental short term stability is mainly due to the fact that the theoretical estimation assumed that the shot noise limit may be reached.
Moreover, there is a difference between the measured and calculated CPT amplitude because the model is based on a pure
bichromatic electromagnetic field, i.e. in the calculation the losses due to the 1-photon absorption of the VCSEL frequencies other than the
first-order sidebands are neglected. The experimental results presented are limited by the laser power: with our VCSEL and the optimal beam diameter (0.8 cm) the maximum laser intensity (I) was 0.1 mW cm$^{-2}$. Following the model, the optimal should be I$=0.4$ mW cm$^{-2}$, corresponding to a shot noise limited short term stability $\approx 1 \cdot 10^{-13} \cdot \tau^{-1/2}$.
\begin{table}[htb]
\centering\caption{Expected short-term stability ($\sigma_A$) in optimized
  conditions.}
\begin{tabular}{lp{1.4in}|lp{1.4in}}
\hline
Parameters & Experiments & Theory\\ \hline
Signal Amplitude, A$_{CPT}$ (nA) & 100 & 400 \\
Signal linewidth, $\Gamma_{CPT}$ (kHz) & 0.7 & 0.7 \\
Detection Noise, N (nA Hz$^{- \frac{1}{2}}$) & 0.03 & 0.005   (Shot Noise)  \\
Short-term stability, $\sigma_A$ & 6 $\cdot$ 10$^{-12}$ $\tau^{- \frac{1}{2}}$& 3 $\cdot$
10$^{-13}$ $\tau^{- \frac{1}{2}}$ \\ \hline
\end{tabular}
\label{tab1}
\end{table}
\section{Conclusion}
We report on our study on the lin$\|$lin CPT for applications in compact and high performance atomic
clocks. We discuss the two relevant cases: degenerate CPT and pseudo-resonance. We showed that a $\sigma_A \sim 6 \cdot$ 10$^{-12}$
at $1$ second may be reached using the degenerate lin$\|$lin CPT signal and a compact setup. This value is comparable to the short-term-stability of commercial compact atomic clocks. On the contrary the short-term stability of a pseudo-resonance clock may be about $2 \cdot$ 10$^{-11}$ at $1$ s.
According to the model, the expected short-term stability is limited by the choice of the laser source and could be improved by using a pure bichromatic fields with higher laser power (up to $0.4$ mW cm$^{-2}$).
\section{Acknowledgement}
The authors are pleased to acknowledge the financial support of the
INTAS-CNES (project 06-1000024-9321), the Swiss National Science
Foundation (project 200020-105624) and the Dimitry Zimin
foundation "Dynasty". Finally we thank for the useful discussion C. Affolderbach, P. Thomann and our collegues of the Laboratoire Temps-Fr\'equence-IMT



\end{document}